\documentclass[global,twocolumn]{svjour}
\usepackage{graphicx}
\journalname{Appl Phys B}

\begin{document}

\title{Bose-Einstein Condensates in Magnetic Waveguides}

\author{J. Fort\'{a}gh, H. Ott, S. Kraft, A. G\"unther, and C. Zimmermann}

\institute{Physikalisches Institut der Universit\"at T\"ubingen\\
 Auf der Morgenstelle 14, 72076 T\"ubingen, Germany}
\date{Received: date / Revised version: date}
\maketitle

\begin{abstract}

In this article, we describe an experimental system for generating
Bose-Einstein condensates and controlling the shape and motion of
the condensate by using miniaturised magnetic potentials. In
particular, we describe the magnetic trap setup, the vacuum
system, the use of dispenser sources for loading a high number of
atoms into the magneto-optical trap, the magnetic transfer of
atoms into the microtrap, and the experimental cycle for
generating Bose-Einstein condensates. We present first results on
outcoupling of condensates into a magnetic waveguide and discuss
influences of the trap surface on the ultracold ensembles.
\end{abstract}

\section{Introduction}

After seven years of research with Bose-Einstein condensates it
was now possible to significantly simplify the complexity of the
first generation experimental apparatus. In recent experiments
condensates have been generated with a duty cycle of only a few
seconds \cite{Barret01,Haensel01}. Furthermore,  it was possible
to avoid Zeeman slowing techniques \cite{Prodan85} or complex
combinations of two magneto-optical traps \cite{Myatt96} by using
a pulsed alkali dispenser source \cite{Fortagh98}. With these
developements, it is now conceivable to construct a compact and
portable apparatus for application oriented experiments with
macroscopic atomic matter waves.  A breakthrough in this direction
has been achieved recently with the successful loading of magnetic
microtraps \cite{Ott01,Haensel01}  with Bose-Einstein condensates.
In principal, magnetic microtraps allow for the construction of a
large variety of trapping potentials. Thus, a growing number of
research groups start to explore the possibilities which arise
from the combination of micropotentials with degenerate atomic
quantum gases \cite{Leanhardt02,Kasper02,Vale02}. Here, we
describe in some detail our experimental approach including the
apparatus that we use to load a condensate into a magnetic
microtrap. First experimental results on outcoupling a condensate
from a trap into the waveguide are presented. We comment recent
measurements of the lifetime, heating rate and fragmentation of
ultracold atomic clouds close to a conductor surface
\cite{Fortagh02cm}. We conclude the article with speculations
about future developments to atom optics on a microchip.

\section{Experimental setup}

For studies on ultracold quantum gases in elongated waveguides, we
have developed a method for loading a large number of atoms into
tightly confining traps at a microstructure. The key feature is
the continuous transformation of a rather shallow magnetic
potential into the tightly confining geometry of the microtrap
\cite{Vuletic98,Fortagh98prl}. The atoms are initially collected
in a magneto-optical trap (MOT) and subsequently loaded into the
shallow magnetic quadrupole trap employing standard techniques of
polarisation gradient cooling and optical pumping. Afterwards, the
atoms are adiabatically compressed into the microtrap by gradually
transforming the magnetic field. The compression enhances the
collision rate and accelerates the thermalisation. When the
collision rate is several 10 per second the atomic gas can be
efficiently cooled by forced evaporation into the degenerate
regime.

\begin{figure} \centering
\includegraphics[width=6.3cm]{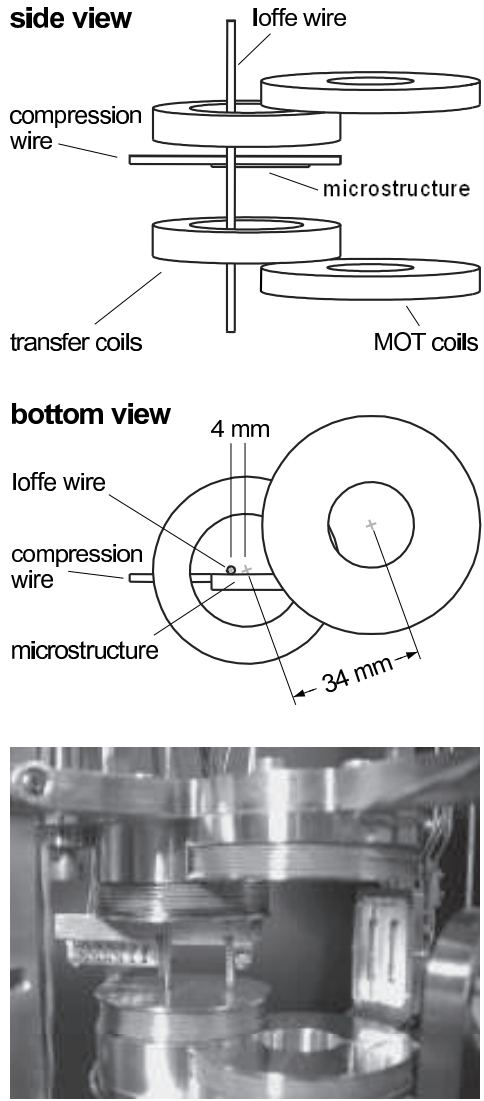}
\caption{\label{setup} Trap setup. The magnetic trap setup
consists of two pair of coils, a Ioffe-wire, a compression wire, a
microstructure with microsized conductors, and a thin copper wire
parallel to the microstructure. The MOT is generated at the right
hand side between the MOT-coils. The magnetic microtrap is
arranged between the transfer coils. The MOT-coils have an inner
diameter of 22 mm and 130 windings on each coil. The distance
between the MOT-coils is 48 mm. The transfer coils have an inner
diameter of 31 mm and 80 windings on each coil. The vertical
distance between the transfer coils is 30 mm. The MOT-coils and
transfer coils are separated by 34 mm. The Ioffe-wire has a
diameter of 2 mm and a circular cross section. It is displaced by
4 mm to the symmetry axis of the transfer coils. The compression
wire has a quadratic cross section with a width of 2 mm. It is
placed horizontally 4 mm above the symmetry plane of the coils and
the centre of the Ioffe-trap \cite{Fortagh00}. The surface of the
microstructure has a distance of 2.2 mm to the central plane of
the coils. The compression wire, the microstructure and the thin
copper wire are mounted on a heat sink between the transfer coils.
The photograph shows the mounted trap setup and the dispenser
sources behind the MOT-coils.  }
\end{figure}
The trap setup employed in our experiments is shown in
Fig.~\ref{setup}. The microtrap is arranged between the \lq
transfer coils\rq . For the operation of the MOT we use a second
pair of coils: the \lq MOT-coils\rq . The MOT is working in a six
beam configuration with beam diameters of 20 mm and 20 mW of laser
power in each beam. The separation of the MOT-coils and transfer
coils guaranties an undisturbed operation of the MOT and allows
for high flexibility in mounting different microtrap geometries.
Thus, different microtraps can be loaded by the same transfer
scheme. The microstructure with microfabricated current conductors
is mounted horizontally upside down below the \lq compression
wire\rq, and the magnetic microtrap is generated below the
microstructure. The Ioffe-geometry of the magnetic traps between
the transfer coils is provided by combining the magnetic fields of
the transfer coils and an additional vertical wire: the \lq
Ioffe-wire\rq\ \cite{Fortagh00}. The Ioffe-wire passes the
transfer coils parallel to their symmetry axis with a lateral
displacement of 4 mm. The Ioffe-trap is built in the middle of a
semi circular trajectory which connects the centre of the
quadrupole field of the transfer coils and the middle of the
Ioffe-wire \cite{Fortagh00}. This trap can be shifted in its
vertical position by varying the currents in the transfer coils.
For the transfer into the microtrap, it is necessary that the
compression wire and at least one of the microfabricated
conductors are oriented parallel to the long axis of the
Ioffe-trap and that they are aligned in the plane in which the
Ioffe-trap can be shifted up and down.

\begin{figure} \centering
\includegraphics[width=5cm]{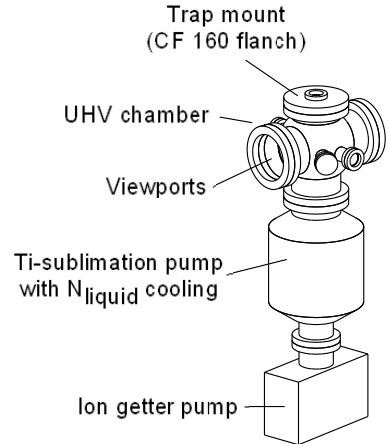}
\caption{\label{vacuum}  The vacuum system. The vacuum chamber is
pumped by a titanium sublimation pump with liquid nitrogen cooling
and an ion getter pump to achieve UHV conditions with a base
pressure of $1\times 10^{-11}$ mbar. The trap setup shown in Fig.1
is mounted at the cover of the vacuum chamber.}
\end{figure}

The trap setup is placed in a vacuum chamber at a base pressure of
$1\times 10^{-11}$ mbar. Ultrahigh vacuum conditions are achieved
in a system shown in Fig.~\ref{vacuum}. The vacuum chamber is
pumped by an ion getter pump (80 l/s) and a titanium sublimation
pump (2500 l/s) cooled by liquid nitrogen. The configuration of
electromagnets and the dispenser sources are mounted at the cover
of the vacuum chamber. View ports with antireflexion coating at
the side and top of the chamber ensure good optical access to the
inside and allow the operation of a six-beam magneto-optical trap.
The \textsl{in-vacuo} setup  is built by UHV compatible components
using OFHC-copper, stainless steel, MACOR,
$\mathrm{Al_2O_3}$-ceramics, ceramic glue, and capton insulated
copper wires. The dissipated heat during the operation of the
microelectromagnets, especially the heat generated by the transfer
coils of about 9 W, by the compression wire and Ioffe-wire of 2 W
each and the microsized conductors of up to 10 W, is conducted by
a copper rod to a liquid nitrogen reservoir outside the chamber.
It ensures a constant temperature for the microstructure at
approximately $-15^{\circ}$C which was determined from the
measured resistance of the copper conductors. Without cooling, the
trap setup warms up, and the lifetime of the atomic clouds is
shortened, nevertheless it can still be used for experiments with
Bose-Einstein condensates.

\begin{figure} \centering
\includegraphics[width=7.5cm]{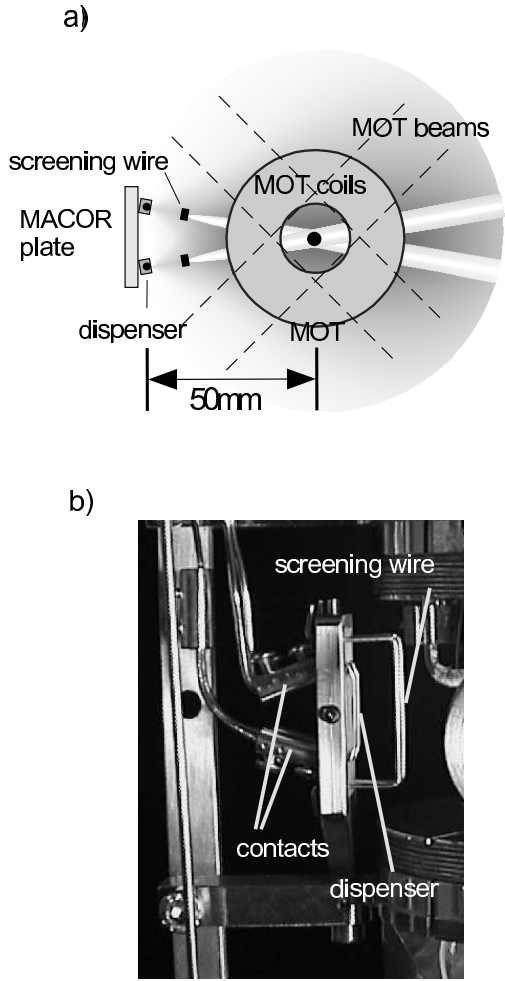}
\caption{\label{dispenser}  The dispenser sources for thermal
rubidium atoms. (a) top view: The MOT is loaded by a pair of
dispensers which emit a beam of rubidium atoms towards the
trapping region of the MOT. The screening wires prevent a direct
impact of the beam to the centre of the MOT. (b) The dispenser
mount (side view). The ends of the dispensers are fed through a
MACOR plate. At the back side, they are fixed to the supplying
copper conductors.}
\end{figure}
As source for thermal rubidium atoms we use a pair of resistively
heated dispensers \cite{Fortagh98}. The dispensers are located at
a distance of 50 mm from the MOT (Fig.~\ref{dispenser}). They are
mounted in a MACOR ceramic plate as shown in
Fig.~\ref{dispenser}b. The ends of the dispensers are bent back at
90 degree and pass through holes in the plate. They  are
mechanically fixed behind the plate to the copper conductors
supplying power. The bending allows for flexing and thermal
expansion of the heated metal container. The dispensers are heated
by  currents of opposite directions to each other. Because  of
this symmetrical configuration it is possible to avoid a shifting
of the MOT position due to the magnetic field of the current flow
in the nearby dispensers. If the dispensers are heated by a
current of a few amps a thermal rubidium beam is emitted into the
trapping region of the MOT (Fig.~\ref{dispenser}a). A screening
wire with 1 mm diameter in front of the rubidium source shields
the centre of the MOT and prevents the magneto-optically trapped
atoms from direct impact of thermal atoms and other contaminations
from the heated area. Without screening, the lifetime of the MOT
is strongly reduced due to fast decay processes on a time scale of
a few seconds. The screening eliminates the losses and a smooth
decay behaviour of the MOT due to the background pressure can be
observed.

\section{The Microstructure}

For the current investigations on ultracold atomic gases, the
magnetic trap is built at a microstructure consisting of seven 22
mm long  parallel copper conductors.
\begin{figure} \centering
\includegraphics[width=6.5cm]{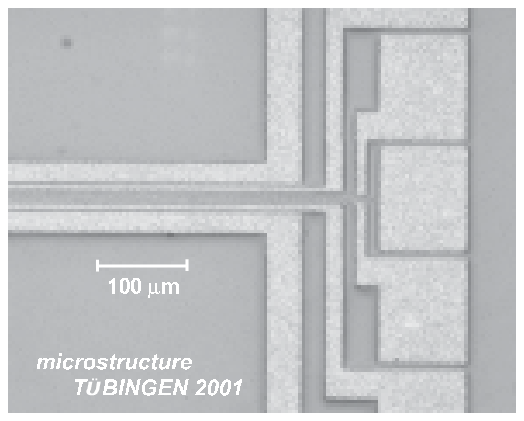}
\caption{\label{microstructure} The microstructure consists of
seven parallel copper conductors with a width of 30, 11, 3, 3, 3,
11, and 30 $\mu$m and a height of $2.5\,\mu$m. On the right hand
side the contact pads are visible (microscope image). }
\end{figure}
The microfabricated conductors are electroplated on an
$\mathrm{Al_2O_3}$ substrate (Fig.~\ref{microstructure}) and have
widths of 30, 11, 3, 3, 3, 11, and 30 $\mu$m, respectively. Their
nominal separation is $1\,\mu$m and the height of the
electroplated copper layer is $2.5\,\mu$m \cite{Fortagh02}. The
microstructure is mounted onto the surface of the compression
wire, with theconductors oriented parallel to the compression
wire. For reference measurements, a thin copper wire with a
diameter of $90\,\mu$m and a length of 25~mm is mounted parallel
to the microstructure. It can be used alternatively to the
conductors at the microstructure for generating a linear microtrap
\cite{Fortagh02cm}. The vertical distance between the middle of
the compression wire and the microfabricated conductors is 1.8 mm
and the distance between the middle of the compression wire and
the thin wire is 2.0 mm.

\begin{figure} \centering
\includegraphics[width=7.5cm]{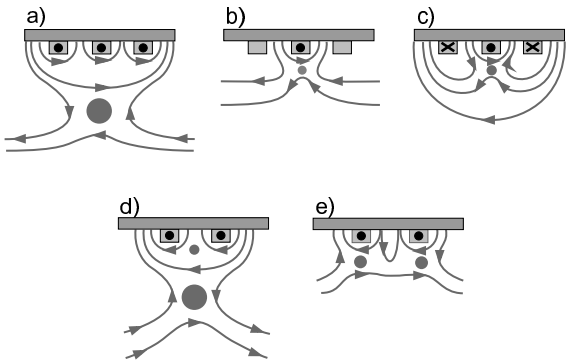}
\caption{\label{traps}  Waveguide geometries at current carrying
linear coductors. The orientation of the driving current is marked
by $\bullet$ and $\times$. The arrows show schematically the slope
of the magnetic field. The filled circles indicates the trap
centres. a) and b) show waveguide geometries formed by the
microstructure and an additional external bias field. In c) the
bias field is generated by the outer conductors of the
microstructure. d) and e) show multiple waveguide configurations.
Depending on the strength of the bias field, the two waveguides
are located side by side or on top of each other.}
\end{figure}

The combination of linear conductors and a bias field
perpendicular to the axis of the conductor enables  a large
variety of radial trap geometries to be built. In the simplest
case, several conductors are driven with currents in the same
direction and the trap is generated by applying an additional bias
field perpendicular to the conductors (Fig.~\ref{traps}a). An
additional axial offset field provides the parabolic potential
shape in the centre of the waveguide.
Turning off the outer conductors shifts the position of the trap
closer to the inner conductor (Fig.~\ref{traps}b) and increases
the radial gradient. The depth of the magnetic trap is determined
by the bias field applied perpendicular to the conductors and does
not change during the compression. The trap can be further
compressed by inverting the currents in the neighbouring
conductors (Fig.~\ref{traps}c). Calculations  with the measured
maximum current densities in the microconductors \cite{Fortagh02}
predict a maximal possible radial oscillation frequency of $2 \pi
\times 600\,000\,$Hz. Starting with standard trap parameters for
Bose-Einstein condensation ($\omega_\mathrm{r} = 2\pi \times
300\,$Hz and $\omega_\mathrm{a} = 2\pi \times 14\,$Hz radial and
axial oscillation frequencies, respectively), it should be
possible to change the aspect ratio of the trap about three orders
of magnitude. A further application of parallel current conductors
is the construction of parallel waveguides \cite{Hinds01}. The
waveguides can be arranged horizontally or vertically to each
other depending on the strength of the applied bias field
(Fig.~\ref{traps}d and ~\ref{traps}e). If the bias field is
oriented exactly parallel to the plane defined by the conductors,
the trajectories of the trap centres cross each other while
changing the bias field and the waveguides can be merged and
separated.

\section{Experimental cycle}

\begin{figure} \centering
\includegraphics[width=7.5cm]{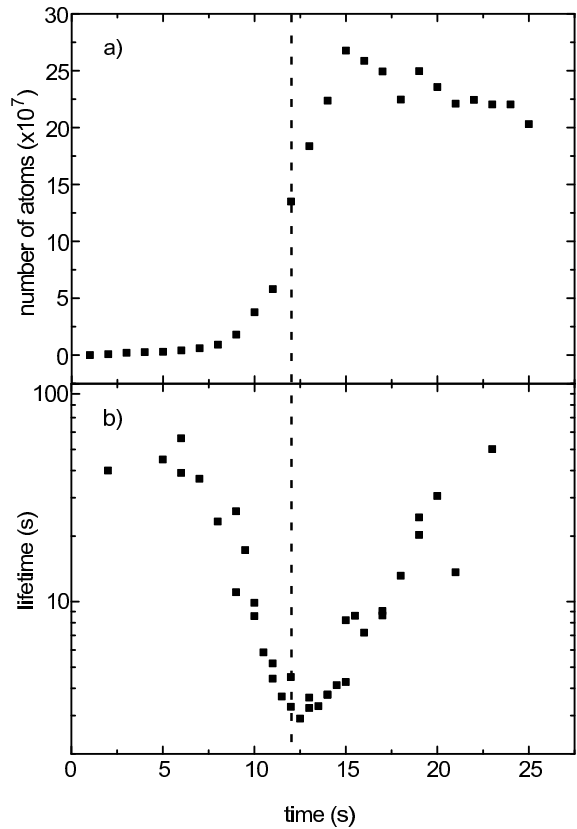}
\caption{\label{mot}  Loading the MOT from the pulsed dispenser
sources. a) Time evolution of the number of atoms in the MOT. b)
Time evolution of the lifetime of trapped atomic clouds between
the MOT-coils during the loading process. The heating current
(6.5A) of the dispenser is turned on at the starting point of the
diagram and it is turned off after 12 s (dashed line). After
switching off the dispenser current, the lifetime recovers in
approximately 10 s after them the atoms are restored into the
magnetic trap.}
\end{figure}

At the begining of the experiment the MOT is loaded within 25 s
from pulsed thermal dispenser sources.  Thereby, $3\times 10^8$
$^{87}$Rb atoms are collected at a temperature of $50\,\mu$K.
During the first 12 seconds, the dispensers are heated by a
constant current pulse of 6.5 A. For the next 13 seconds, the
dispensers are turned off and the MOT is loaded from the thermal
rubidium beam and the residual vapour. The time evolution of the
number of atoms and their lifetime is shown in Fig.~\ref{mot}.
Approximately seven seconds after switching on the heating
current, the dispensers reach the dissociation temperature of
rubidium and the emission starts. The local pressure of rubidium
grows and the MOT is loaded on a time scale within its lifetime.
After switching off the heating, the dispensers rapidly cool down
below the dissociation temperature,
predominantly due to thermal radiation \cite{Fortagh98}. Within
the next few seconds, the rubidium vapour is pumped out and the
initial trap lifetime of about 60 s is recovered. The data points
in Fig.~\ref{mot}a and Fig.~\ref{mot}b have been measured in
separate experimental cycles. The number of atoms was detected by
absorption imaging.

The preparation of ultracold atoms in the MOT is finished by 10 ms
polarisation gradient cooling. Then, the atoms are optically
pumped into the $|\mathrm{F}\!\!=\!\!2, \mathrm{m_F}\!\!=\!\!2\!>$
hyperfine ground state and a $60\,\mu$K cold cloud of $2\times
10^8$ atoms is loaded into the spherical quadrupole trap formed by
the MOT-coils at a field gradient of 45 G/cm. Because the
MOT-coils and the transfer coils overlap, an adiabatic transfer of
the magnetically trapped atoms can be performed: the quadrupole
potential minimum moves on a straight line from the centre of the
MOT-coils to the centre of the transfer coils. At a current of 3 A
the transfer coils generate a spherical quadrupole field with a
gradient of 58 G/cm along the symmetry axis. By increasing the
current in the Ioffe-wire, the centre of the spherical quadrupole
is shifted and transformed into a Ioffe-type trapping field
\cite{Fortagh00}.  At a current of 13 A in the Ioffe-wire the
resulting harmonic trap potential is characterised by its axial
oscillation frequency of $\omega_\mathrm{a} = 2\pi \times 14\,$Hz,
its radial oscillation frequency of $\omega_\mathrm{r}= 2\pi\times
110\,$Hz, and the offset field of 0.4 G in the middle of the trap.
In this Ioffe-type trap the atoms are cooled for 20 s by radio
frequency evaporation to a temperature of $5\,\mu$K. In this large
volume Ioffe-trap, condensation can be reached with approximately
$2\times 10^4$ atoms if the cloud is cooled for a further 10 s.
However, the transfer into the microtrap offers more advantageous
conditions for condensation.

\begin{figure} \centering
\includegraphics[width=7.5cm]{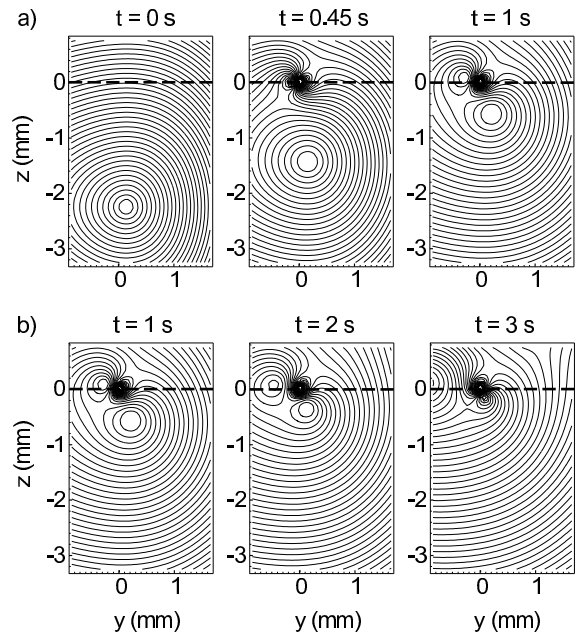}
\caption{\label{transfer}  Magnetic field contour plots (contour spacing 0.5
  G) in the plane perpendicular to the microsized conductor located at y=0,
  z=0. The dashed line indicates the position of the microstructure's
  surface. The large volume Ioffe-trap is located at y=0, z=-2.2, its long
  axis is perpendicular to the illustrated plane.
a) Transfer from the large volume Ioffe-trap into the microtrap within 1
   s. The current in the microconductor (0.4 A) is switched on and the
   Ioffe-trap is shifted up by lowering the current in the upper transfer coil
   from 3 A to 2.3 A. The lower transfer coil carries a current of 3 A, the
   Ioffe-wire 13 A. As a result, the trap shifts up by approximately 1.6 mm
   and ends at a vertical displacement of $600\,\mu$m to the microsized
   conductor. b) Compression of the microtrap within 2 s. The current in the
   compression wire is increased to 2.7 A while simultaneously lowering the
   current in the upper transfer coil to 2 A. Reducing the current in the
   transfer coil is needed to avoid the saddle point moving into the
   microtrap. After the shown compression, the trap is located at a vertical distance
   of $200\,\mu$m to the middle of the microconductor.
   }
\end{figure}

Therefore, after $20\,$s of precooling in the large volume
Ioffe-trap, the ensemble is compressed adiabatically into the
microtrap by reducing the current in the upper transfer coil. This
shifts the trap minimum to the microstructure
(Fig.~\ref{transfer}). For a tighter confinement of the cloud in
the microtrap, a current in the compression wire can be applied.
Good conditions for Bose-Einstein condensation have been achieved
for a large range of radial oscillation frequencies between
$\omega_\mathrm{r} = 2\pi \times 200\,$Hz and $2\pi \times
2\,000\,$Hz. The axial confinement is determined by the Ioffe-trap
to $\omega_\mathrm{a} = 2\pi\times 14\,$Hz in all experiments. The
compression from the Ioffe-trap into the microtrap takes place
within 1 s and is completed by applying the current in the
compression wire. During the transfer and compression the radio
frequency is turned off. The compression heats the atomic cloud by
a factor of up to 7 and boosts the elastic collision rate to
several hundred per second. By now ramping down the radio
frequency from 5 MHz to about 1 MHz within 5 s we reach
condensation with up to $1\times 10^6$ atoms at a critical
temperature of typically between 500 nK and $1\,\mu$K
\cite{Ott01}.

The radial confinement of an elongated trap is determined in
principal by the current in the microconductor and the compression
wire \cite{Fortagh02cm}. However, additional bias fields from the
Ioffe-wire and the transfer coils can significantly modify this
value. Therefore, we measure the exact trap frequencies by
sinusoidal perturbating the trapping potential and observing the
temperature of the thermal cloud. To complete the characterisation
of the trap, we measure the bias field in the centre of the trap
by removing all the atoms via rf-outcoupling.

The parameters of the Bose-Einstein condensate such as the number
of atoms, the chemical potential and the density are derived from
absorption images after several milliseconds time of flight.
Condensates in tightly confining traps are generated at high
densities of about $10^{15}\,\mathrm{cm}^{-3}$. Due to inelastic
three body losses \cite{Burt97} the lifetime is in the order of a
hundred millisecond. By reducing the density, the lifetime of the
condensate can be enhanced up to several seconds in which
subsequent investigations on the Bose-Einstein condensate can be
performed \cite{Fortagh02cm}. Thereby, magnetic microtraps allow a
fast and versatile manipulation of the quantum gas. Altogether,
the experimental cycle requires approximately one minute.

\section{Experiments}

In applications on microtraps, the propagation of condensates in waveguides is of
particular interest. An atomic beam could be guided from an on-chip atom laser to an
experimental facility. For this reason an understanding of the outcoupling process from a trap into
a waveguide and the physical properties of propagating condensates is required. We have
started to study experimental scenarios where the condensate is released into a waveguide.
Beginning with a condensate in a trap, characterized by the radial and axial oscillation
frequencies of $\omega_\mathrm{r} = 2\pi  \times 500\,$Hz and $\omega_a =2\pi \times 14\,$Hz,
respectively, the axial
confinement was turned off linearly within 0.4 s. An additional gradient
field was turned on parallel to the guide axis which forces the condensate into the waveguide.
\begin{figure} \centering
\includegraphics[width=7.5cm]{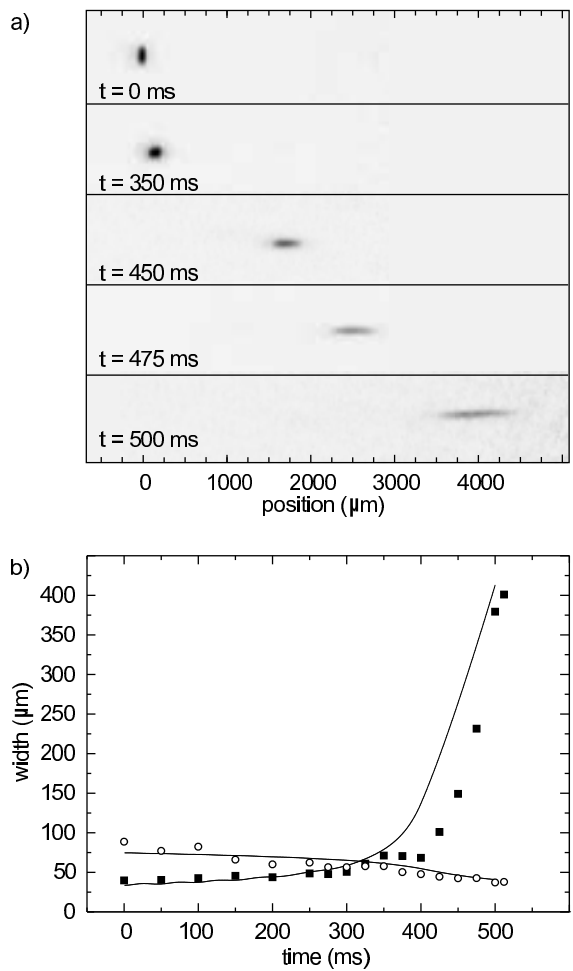}
\caption{\label{waveguide}  Expansion of a condensate into the
waveguide. The axial confinement of the magnetic trap is turned
off within 400 ms and a gradient field forces the condensate into
the waveguide. The release of the condensate into the waveguide is
completed after 385 ms.
 (a)  Position of the condensate at different stages of the expansion, after
 23 ms time of flight (absorption images). The initial trap is located at the
 origin. (b) Time evolution of the radial (open circles) and the axial (filled
 squares) size of the condensate during the expansion. The straight lines show
 a numerical simulation of the time evolution of the condensate (see text).}
\end{figure}
In Fig.~\ref{waveguide}a five steps of the process are shown: the
first image shows the condensate released from the initial trap
after 23 ms time of flight. The condensate expands mostly in
radial direction according to the tight confinement and high
interaction energy. 0.35 s later, the axial confinement is already
strongly reduced and the smaller radial size of the expanded
condensate in the time of flight image indicates a decreased
interaction energie. After 0.35 s the axial opening of the trap is
no longer adiabatic due to the extreme low axial trap frequencies.
15 ms before the axial confinement is completely turned off, the
condensate is released from the initial trap ($t = 385\,$ms). The
third image shows the condensate after 65 ms of propagation in the
waveguide. The condensate still expands axially. Due to the
external gradient field, it is accelerated by
$0.38\,\mathrm{m/s^2}$. After 125 ms of propagation, it reaches a
velocity of 48 mm/s. Fig.~\ref{waveguide}b shows the evolution of
the expanded radial and axial width of the condensate after 23 ms
time of flight during the expansion process. The radial width of
the condensate decreases as the axial expansion lowers the density
and the chemical potential. The straight lines in
Fig.~\ref{waveguide}b are a numerical simulation of the radial and
axial halflenghts of the condensate. The simulation is based on
the scaling equations for a condensate in a time-dependent
harmonic trap \cite{Castin96,Kagan97}. The trap was assumed to be
at rest, while the axial curvature was linearly ramped down within
0.4 s. The centre of mass motion is decoupled from the inner
excitations of the condensate which are well described by the
theory. Note, that there is no fitting parameter in the theory. In
future experiments, a detailed study of dynamics and coherence
properties of elongated condensates and propagating beams may
become feasible.

\begin{figure} \centering
\includegraphics[width=7.5cm]{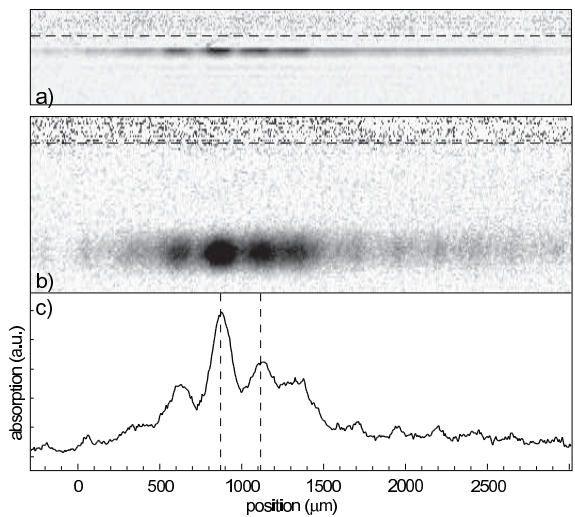}
\caption{\label{splitting} The spatial density modulation of an
expanding atomic cloud in the waveguide indicates a periodic
potential structure along the current carrying copper conductor.
For the experiment, an ultracold cloud of $5.6\times 10^5$ atoms
at the temperature of $1\, \mu$K was released into the waveguide
by turning off the axial confinement of the trap within 400 ms.
The waveguide had a radial confinement of $\omega_\mathrm{r} =
2\pi \times 1000\,$Hz. The absorption images were taken a) in the
waveguide and b) after 10 ms time of flight. c) The integrated
scan shows a periodicity of $260\, \mu$m in the density
distribution of the cloud. }
\end{figure}

In recent experiments, the heating rate and the lifetime have been
investigated on atomic clouds at small distances ($20 - 300\,
\mu$m) to the surface of the microtrap \cite{Fortagh02cm}. It has
been shown that the heating rate remains at each distances below
400 nK/s whereas the lifetime experience a reduction when
approaching the surface. At a distance of $200\, \mu$m we measured
a lifetime of 13 s which is reduced almost linearly to 2 s when
the distance was reduced to $30\, \mu$m. However, the most
peculiar effect which appears if atoms are moved into traps close
to the surface of the conductor, is a fragmentation of the cloud
along the conductor \cite{Fortagh02cm}. The fragmentation can be
observed on ultracold and on condensed clouds. We have made the
observations at the thin copper wire as well as along each
microfabricated conductor. Fig.~\ref{splitting} shows the
fragmentation of an expanding atomic cloud in a waveguide. These
experiments indicate the general appearance of a periodic
potential structure along current carrying copper conductors. The
most probable structuring potential type is a magnetic one.
Because an explanation of having a periodic structure of
ferromagnetic contaminations on the conductors of different origin
seems to be unlikely, we speculate whether the modulation
corresponds to spin arrangements of moving electrons, a topic
which is investigated be the research field of spintronics
\cite{Sarma99}. Meanwhile, the observations discussed above could
be reproduced by two other groups \cite{Leanhardt02,Vale02}
working on microtraps with copper conductors.

\section{Outlook}

Trapping and manipulating atoms in magnetic microtraps is a very
young field and it is difficult to make predictions about future
developments. Nevertheless, one can speculate about future
research topics. The experiments presented in \cite{Fortagh02cm}
show a strong influence of a nearby surface on a condensate. This
suggests the use of  the condensate as a probe to study an
arbitrary surface. The sample surface can be either metallic or
dielectric or a combination of both. Electric or magnetic surface
forces, acting on the condensate will structure its density
distribution. This can be observed with a variety of well
established imaging techniques. In this case the resolution is
limited by the imaging system to the micron range. A better
resolution may be achieved by imaging the condensate after some
milliseconds of free expansion in the gravitational field. Since
such \lq time of flight images\rq\ reflect the initial velocity
distribution of the condensates, they would provide information
about the phase gradients inside the condensate as imprinted by
the surface under investigation. Such an approach can be regarded
as  an extension of recent experiments with condensates in three
dimensional optical lattices \cite{Greiner01}. By adiabatically
reducing the lattice potential followed by a ballistic expansion,
it was possible to directly image the population of the energy
bands within the first Brillouin-zone.  The size of the
Brillouin-zone increases with decreasing lattice constant which
means that the size of the time of flight images increases with
decreasing size of the periodic structure. The limited resolution
of the imaging system can be overcome in this way. By decomposing
an arbitrary surface potential into periodic Fourier components
one may be able to develop a novel kind of holographic surface
microscope. These experiments can readily be done in a
conventional Bose-Einstein apparatus by mounting a sample surface
close to the centre of the magnetic trap and shifting the
condensate towards the surface by varying the magnetic field of
the trap in an appropriate way.

Another fascinating topic for future research with microtraps is
the development of elements for integrated atom optical devices.
If condensates can be split and recombined with temporal or
spatial beam splitters it is conceivable to construct
interferometers that are sensitive to rotations and forces
\cite{review}. Micropotentials can also be used to control the
relative position between different atomic clouds or even single
atoms. This allows for a tunable interaction between atoms and can
be used for novel approaches to construct quantum
gates\cite{Jaksch99}.

In general, it appears that the generic dynamical process in
magnetic micro potentials would be an oscillation  rather than a
current as in electronic circuits. Functional units may thus be
envisioned as being constructed from oscillations that  are
conditionally coupled via their phase or amplitude. Quantum
computers and other \lq atom chip\rq\ devices may turn out to be a
sophisticated combination of such \lq computons\rq . In our
experiment we have observed centre of mass oscillations of
condensates with a very large quality factor. After suddenly
shifting the minimum of the trapping potential that contains a
condensate the resulting sloshing mode can be observed for several
seconds. Its frequency of 7.9 Hz is determined with an relative
error of $7 \times 10^{-4}$ by fitting the data to a sine
function. Although half of the atoms are lost during the
experiment the oscillation amplitude is undamped within the
uncertainty of the measurement and  we estimate the quality factor
of the oscillation  to be better than $10\,000$.  Magnetic
microtrap potentials  can be shaped in almost arbitrary ways such
that nonlinear oscillations can be easily studied. In such
potentials the centre of mass motion is coupled to collective
excitations of the condensate and provides a periodic drive for
the condensate shape oscillation. For the future it will certainly
be interesting to investigate the dynamics of such a nonlinear
system and study its suitability for the construction of atomic
devices.

\section{Acknowledgements}

We thank G.~Schlotterbeck, B.~Herzog, and D. Wha\-ram for the
production of the microstructure, G.~Ritt and M.~Ruder for the
imaging software, S.~G\"unther and C.~Silber for numerical
calculations, L. Borda for calculations on 1D systems, and
G.~Mihaly for discussions about spintronics. This work was
supported in part by the Deutsche Forschungsgemeinschaft under
Grant No.~Zi 419/3.


\begin{thebibliography}{99}
\bibitem{Barret01}
M. D. Barrett, J.A. Sauer, and M.S. Chapman, Phys. Rev. Lett. \textbf{87,} 010404 (2001)
\bibitem{Haensel01}
W. H\"ansel, P. Hommelhoff, T.W. H\"ansch, and J. Reichel, Nature
\textbf{413,} 498 (2001)
\bibitem{Prodan85}
J. Prodan, A. Migdall, W. D. Phillips, I. So, H. Metcalf, and J. Dalibard,
Phys. Rev. Lett. \textbf{54,} 992 (1985)
\bibitem{Myatt96}
C. J. Myatt, N. R. Newbury, R. W. Ghrist, S. Loutzenhiser, and C. E. Wieman,
Opt. Lett. \textbf{21,} 290 (1996)
\bibitem{Fortagh98}
J. Fortagh, A. Grossmann, T.W. H\"ansch, and C. Zimmermann,
J. Appl. Phys. \textbf{84,} 6499 (1998)
\bibitem{Ott01}
H. Ott, J. Fortagh, G. Schlotterbeck, A. Grossmann, and C. Zimmermann,
Phys. Rev. Lett. \textbf{87,} 230401 (2001)
\bibitem{Leanhardt02}
A. E. Leanhardt, A. P. Chikkatur, D. Kielpinski, Y. Shin, T. L. Gustavson,
W. Ketterle, and D. E. Pritchard, Phys. Rev. Lett. \textbf{89,} 040401 (2002)
\bibitem{Kasper02}
A. Kasper, S. Scheider, C. v. Hagen, L. Feenstre, J. Schmiedmayer, 18th International Conference on Atomic Physics, Boston (2002)
\bibitem{Vale02}
M. P. A. Jones, C. J. Vale, K. Furusawa, E.A. Hinds, 18th International Conference on Atomic Physics, Boston (2002)
\bibitem{Fortagh02cm}
J. Fortagh, H. Ott, S. Kraft, and C. Zimmermann, cond-mat/0205310 (2002)
\bibitem{Vuletic98}
V. Vuletic, T. Fischer, M. Praeger, T. W. H\"ansch, and C. Zimmermann,
Phys. Rev. Lett. \textbf{80,} 1634 (1998)
\bibitem{Fortagh98prl}
J. Fortagh, A. Grossmann, C. Zimmermann, and T. W. H\"ansch
Phys. Rev. Lett. \textbf{81,} 5310 (1998)
\bibitem{Fortagh00}
J. Fortagh, H. Ott, A. Grossmann, and C. Zimmermann, Appl. Phys. B
 \textbf{70,} 701 (2000)
\bibitem{Fortagh02}
J. Fortagh, H. Ott, G. Schlotterbeck, C. Zimmermann, B. Herzog, and D. Wharam,
Apl. Phys. Lett. \textbf{81,} 1146 (2002)
\bibitem{Hinds01}
E. A. Hinds, C. J. Vale, and M. G. Boshier, Phys. Rev. Lett. \textbf{86,} 1462
(2001)
\bibitem{Burt97}
E. A. Burt,  R. W. Ghrist, C. J. Myatt, M. J. Holland, E. A. Cornell, and
C. E. Wieman, Phys. Rev. Lett. \textbf{79,} 337 (1997)
\bibitem{Castin96}
Y. Castin and R. Dum, Phys. Rev. Lett. \textbf{77,} 5315 (1996)
\bibitem{Kagan97}
Yu. Kagan, E. L. Surkov, and G. V. Shlyapnikov, Phys. Rev. A \textbf{55,} R18 (1997)
\bibitem{Sarma99}
S. Das Sarma, J. Fabian, X. Hu, I. \v{Z}uti\'{c}, cond-mat/9912040 (2000)
\bibitem{Greiner01}
M. Greiner, I. Bloch, O. Mandel, T. W. H\"ansch, and T. Esslinger,
Phys. Rev. Lett. \textbf{87,} 160405 (2001)
\bibitem{review}
J. Reichel, W. H\"ansel, P. Hommelhoff, and T. W. H\"ansch, Appl. Phys. B
\textbf{72,} 81 (2001),
R. Folman, P. Kr\"uger, J. Schmiedmayer, J. Denschlag, and C. Henkel,
to appear in Adv. Opt. Mol. Phys. Vol 48 (Academic, New York 2002)
\bibitem{Jaksch99}
D. Jaksch, H.-J. Briegel, J. I. Cirac, C. W. Gardiner, and P. Zoller,
Phys. Rev. Lett. \textbf{82,} 1975 (1999)

\end{thebibliography}
%

\end{document}